\documentclass[12pt,twocolumn,a4paper]{article}

\usepackage{times}
\usepackage{ucs}
\usepackage[utf8x]{inputenc}
\usepackage{graphicx}
\usepackage{amsmath,url}

\newcommand{\dd}{{\text{d}}}
\newcommand{\TT}{{\mathcal{T}}}
\newcommand{\Ol}{\mathcal{O}}

\title{Atomic clocks: new prospects in metrology and geodesy}
\author{Pac\^ome Delva\thanks{Corresponding author. E-mail:
Pacome.Delva@obspm.fr}, J\'er\^ome Lodewyck 
\\ \\ \small{LNE-SYRTE, Observatoire de Paris, CNRS, UPMC ; 61 avenue de
l'Observatoire, 75014 Paris, France} }

\begin{document}

\maketitle

\begin{abstract}
We present the latest developments in the field of atomic clocks and their applications in metrology and fundamental physics. In the light of recent advents in the accuracy of optical clocks, we present an introduction to the relativistic modelization of frequency transfer and a detailed review of chronometric geodesy.
\end{abstract}

\begin{figure}
	\includegraphics[width=\columnwidth]{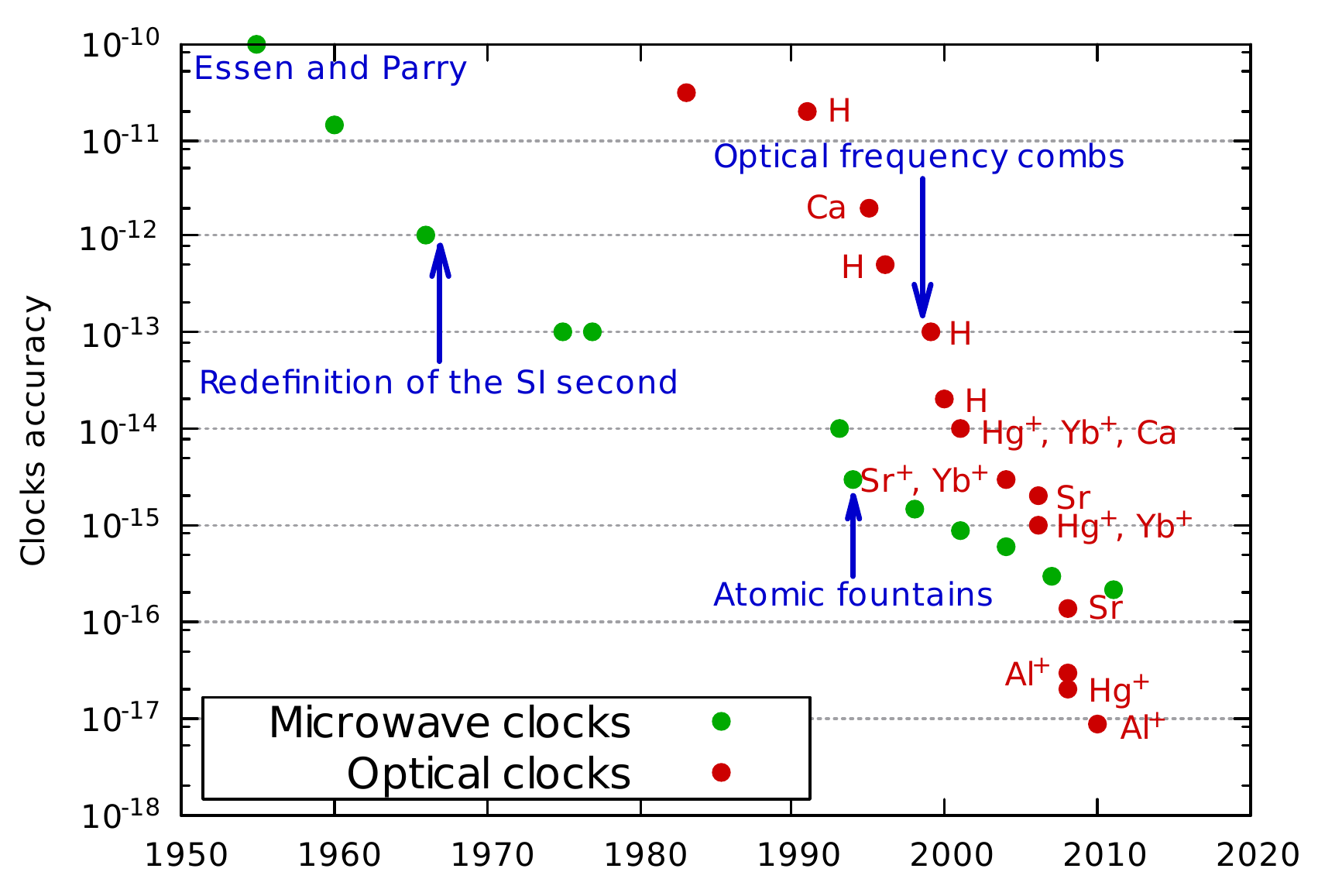}
	\caption{\label{fig:accuracy}Accuracy records for microwave and optical clocks. From the first Cs clock by Essen and Parry in the 1950's, an order of magnitude is gained every ten years. The advent of optical frequency combs boosted the performances of optical clocks, and they recently overcome microwave clocks.}
\end{figure}

\section{Introduction}
Atomic clocks went through tremendous evolutions and ameliorations since their
invention in the middle of the twentieth century. The constant amelioration of
their accuracy (figure~\ref{fig:accuracy}) and stability permitted numerous applications in the field of
metrology and fundamental physics. For a long time cold atom Caesium fountain
clocks remained unchallenged in terms of accuracy and stability. However this is
no longer true with the recent development of optical clocks. This new generation of
atomic clock opens new possibilities for applications, such as chronometric
geodesy, and requires new developments, particularly in the field of frequency
transfer. The LNE-SYRTE laboratory (CNRS/LNE/Paris Observatory/UPMC) is involved in
many aspects of the development of atomic clocks and their applications.

In section~\ref{sec:atcl} we present the latest developments in the field of
atomic clocks: microwave clocks, optical clocks, their relation to international
time-scales, means of comparisons and applications. Section~\ref{sec:relat} is
an introduction to relativistic time transfer, the modelization of remote
frequency comparisons, which lies at the heart of many applications of atomic
clocks, such as the realization of international time-scales and chronometric
geodesy. Finally, section~\ref{sec:chrogeo} is a detailed review of the field of
chronometric geodesy, an old idea which could become reality in the near future.

	\section{Atomic clocks}
\label{sec:atcl}

		In 1967, the definition of the SI second was changed from astronomical references to atomic references by setting the frequency of an hyperfine transition in the Cs atom~\cite{0026-1394-4-1-006}. Since then, the accuracy of atomic clocks has improved by five orders of magnitude, enabling better and better time-keeping. More recently, a new generation of atomic clocks, based on atomic transitions in the optical domain are challenging the well established Cs standard and thus offer opportunities for new applications in fundamental physics and geodesy.

	\subsection{Microwave clocks}

		In a microwave atomic frequency standard, a microwave electro-magnetic radiation excites an hyperfine electronic transition in the ground state of an atomic species. Observing the fraction of excited atoms $p$ after this interaction (or transition probability) gives an indicator of the difference between the frequency $\nu$ of the microwave radiation and the frequency $\nu_0$ of the hyperfine atomic transition. This frequency difference $\nu - \nu_0$ (or error signal) is fed in a servo-loop that keeps the microwave radiation resonant with the atomic transition. According to Fourier's relation, the frequency resolution that can be achieved after such an interrogation procedure grows as the inverse of the interaction time $T$, and since consecutive interrogations are uncorrelated, the frequency resolution further improves as the square root of the total integration time $\tau$. Quantitatively, the residual frequency fluctuation of the microwave radiation locked on the atomic resonance are (in dimension-less fractional units, that is to say divided by the microwave frequency):
		\begin{equation}
			\label{eq:stab}
			\sigma_y(\tau) = \frac{\xi}{\nu_0 T \sqrt{N}}\sqrt{\frac{T_c}{\tau}},
		\end{equation}
		where $T_c$ is the cycle time (such that $\tau/T_c$ is the number of clock interrogations), and $N$ is the number of simultaneously (and independently) interrogated atoms. $\xi$ is a numerical constant, close to unity, that depends on the physics of the interaction between the radiation and the atoms. This expression is the ultimate \emph{frequency (in)stability} of an atomic clock, also called the \emph{Quantum Projection Noise} (QPN) limit, referring to the quantum nature of the interaction between the radiation and the atoms. It is eventually reached if all other sources of noise in the servo-loop are made negligible.

		As seen from eq.~(\ref{eq:stab}), an efficient way to improve the clock stability is to increase the interaction time $T$. The first atomic clocks therefore comprised a long tube in which a thermal beam of Cs atoms is travelling while interacting with the microwave radiation.

		\begin{figure}
			\begin{center}
				\includegraphics[width=\columnwidth]{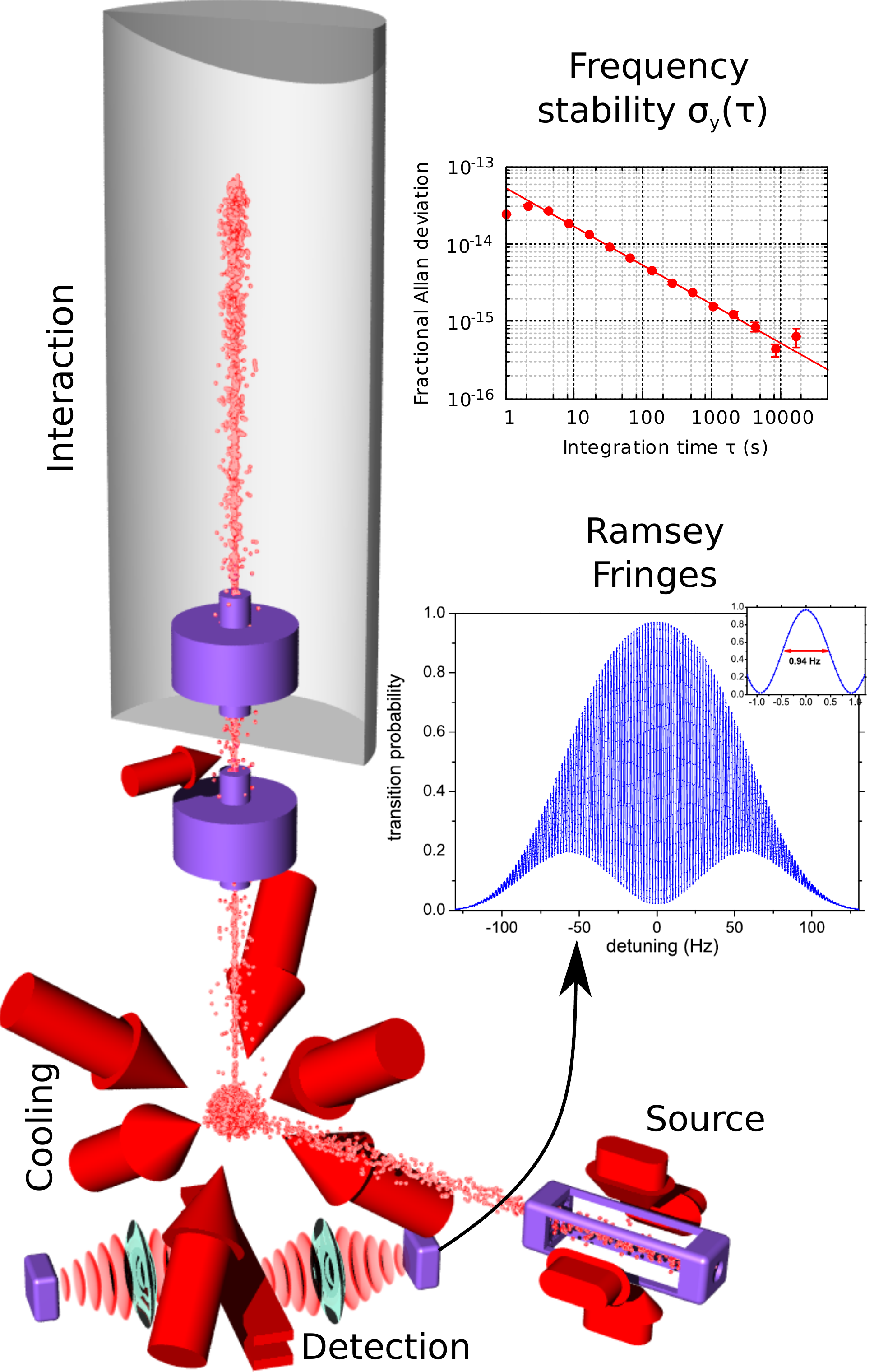}
			\end{center}
			\caption{\label{fig:fountain}Atomic fountain clock. About $10^6$ cold atoms, at a temperature of 1~$\mu$K are launched upwards by pushing laser beams. At the beginning and the end of their trajectory, they interact with a microwave radiation in a resonator and the hyperfine transition probability is measured by an optical detection. When scanning the microwave frequency, the transition probability follows a fringe (Ramsey) pattern much like a double slit interference pattern. During the clock operation, the microwave frequency is locked on the top of the central fringe. The frequency stability, defined by eq.~(\ref{eq:stab}), is a few $10^{-14}$ after 1~s, and a statistical resolution down to $10^{-16}$ is reached after a few days of continuous operation~\cite{6174184,0026-1394-47-1-001}.}
		\end{figure}

		The advance in the physics of cold atoms enabled to prepare atoms with a smaller velocity and consequently largely increase the interaction time, thus the frequency stability of atomic standards. In cold atoms clocks, the interrogation time is only limited by the time after which the atoms, exposed to gravity, escape the interrogation zone. In the atomic fountain clock~(see fig.~\ref{fig:fountain}), a set of cold atoms are launched vertically in the interaction zone and are interrogated during their parabolic flight in the Earth gravitational field.

		In micro-gravity, atoms are not subject to the Earth gravitational field
        and thus can be interrogated during a longer time window $T$. This will
        be the case for the Pharao space microwave clock~\cite{ACES2009}.

		The reproducibility of an atomic clock stems from the fact that all atoms of a given species are rigorously indistinguishable. For instance, two independent atomic clocks based on caesium will produce the same frequency (9,192,631,770 Hz exactly in the SI unit system), regardless of their manufacturer or their location in space-time. 

		However, in laboratory conditions, atoms are surrounded by an experimental environment that perturbs their electronic state and thus slightly modify their resonance frequency in a way that depends on experimental conditions. For instance, atoms will interact with DC or AC electro-magnetic perturbations, will be sensitive to collisions between them in a way that depends on the atomic density or to the Doppler effect resulting from their residual motion\ldots Such systematic effects, if not evaluated and correctd for, limit the universality of the atomic standard. Consequently, the accuracy of a clock quantifies the uncertainty on these systematic effects. Currently, the best microwave atomic fountain clocks have a relative accuracy of $2\times10^{-16}$, which is presumably their ultimate performances given the many technical obstacles to further improve this accuracy.

	\subsection{SI second and time scales}

About 250 clocks 
worldwide, connected by time and frequency transfer techniques (mainly through
GNSS signals) realize a complete architecture that enable the creation of atomic
time scales. For this, local time scales physically generated by metrology
laboratories are \emph{a posteriori} compared and common time-scales are decided
upon. the International Bureau for Weights and Measures (BIPM) in Sèvres
(France) is responsible for establishing the International Atomic Time (TAI).
First a free atomic time-scale is build, the EAL. However, this time-scale is
free-running and the participating clocks do not aim at realizing the SI second.
The rate of EAL is measured by comparison with a few number of caesium atomic
fountains which aim at realizing the SI second, and TAI is then derived from EAL
by applying a rate correction, so that the scale unit of TAI is the SI second as
realized on the rotating geoid~\cite{CCDS1980}. It necessary to take into
account the atomic fontain clocks frequency shift due to relativity (see
section~\ref{sec:relat}). Recently, a change of paradigm occured in the
definition of TAI. According to UAI resolutions~\cite{soffel03}, TAI is a realization of
Terrestrial Time (TT), which is defined by applying a constant rate correction
to Geocentric Coordinate Time (TCG). This definition has been adopted so that
the reference surface of TAI is no longer the geoid, which is not a stable
surface (see section~\ref{sec:chrogeo}). Finally, the Coordinated Universal Time
(UTC) differs from TAI by an integer number of second in order to follow the
irregularities of the Earth rotation.


The publication of such time-scales enables world-wide comparison of Cs fountain
clocks~\cite{0026-1394-47-1-001, 6377298} and the realization of the SI second.

	\subsection{Optical clocks}

		\begin{figure}
			\begin{center}
				\includegraphics[width=\columnwidth]{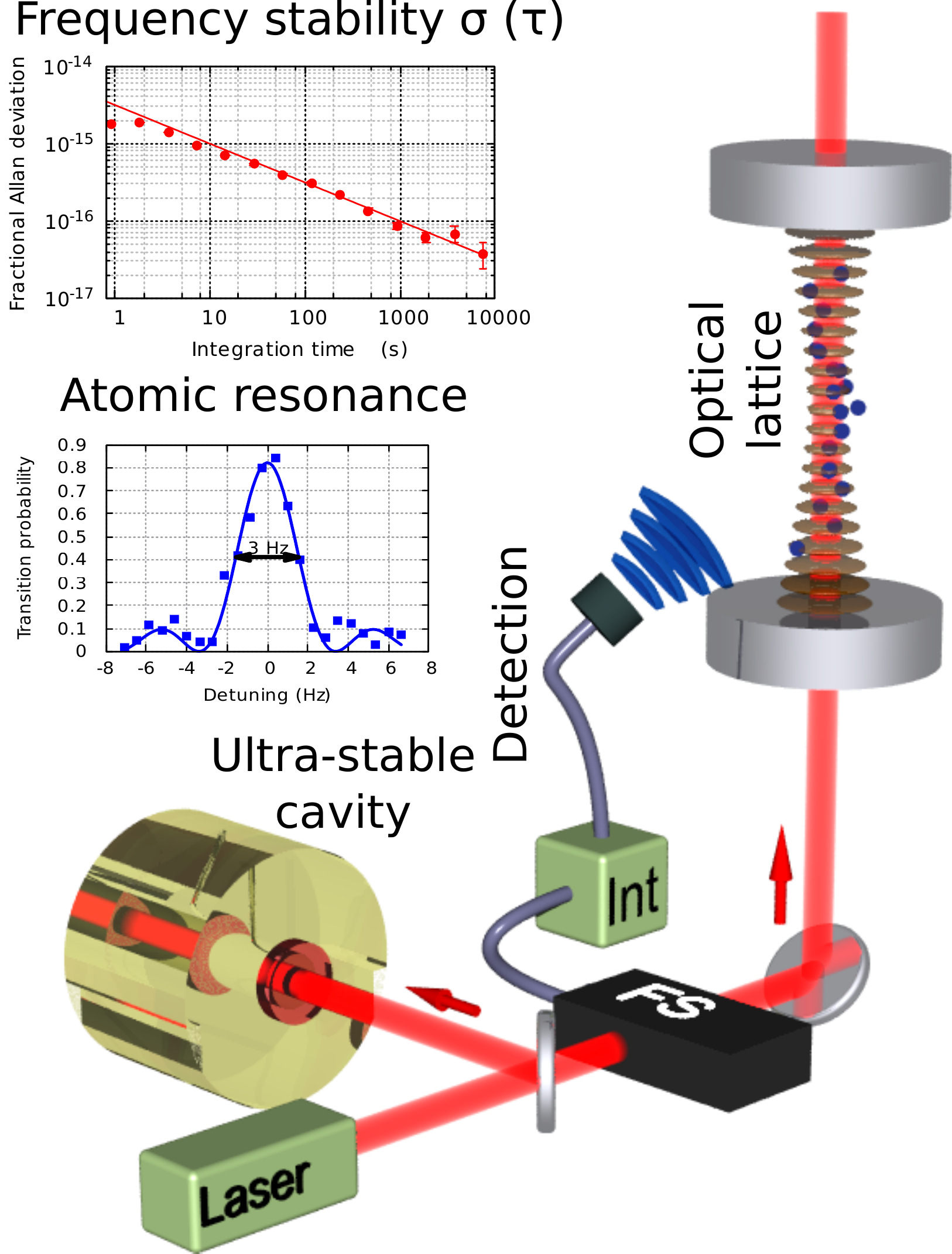}
			\end{center}
			\caption{\label{fig:sr}Sr optical lattice clock. The clock laser,
            prestabilized on an ultra-stable optical cavity, probes an optical
            transition of $10^4$ atoms confined in an optical lattice. The
            excitation fraction (or transition probability) is detected by a
            fluorescence imaging and a numerical integrator acts on a frequency
            shifter (FS) to keep the laser on resonance with the atomic
            transition. The width of the resonance is Fourier-limited at 3~Hz, which, given the clock frequency $\nu_0 = 429$~THz, yields a resonance quality factor $ Q = 1.4\times10^{14}$, compared to $10^{10}$ for microwave clocks.}
		\end{figure}

		A new generation of atomic clocks have appeared in the last 15 years. These clocks consist in locking an electromagnetic radiation in the optical domain ($\nu = 300$ to 800~THz) to a narrow electronic transition. As seen from eq.~(\ref{eq:stab}), increasing the frequency  $\nu_0$ of the clock frequency by several orders of magnitude drastically improves the ultimate clock stability, even though the number of interrogated atoms $N$ is usually smaller in optical clocks. Increasing the clock frequency also improves the clock accuracy since most systematic effects (sensitivity to DC electromagnetic fields, cold collisions\ldots) are of the same order of magnitude in frequency units, and thus decrease in relative units. However, two notable exceptions remain, and both have triggered recent research in optical clocks. First, the sensitivity of the clock transition frequency on the ambient black-body radiation is mostly rejected in microwave clocks but not for optical transitions. Thus, the uncertainty on this effect usually only marginally improves when going to optical clocks (with the notable exception of a few atomic species such as Al$^{+}$ for which the sensitivity is accidentally small). Therefore, extra care has to be taken to control the temperature of the atoms environment. Second, the Doppler frequency shift $\delta\nu$ due to the residual velocity $v$ of the atoms scales as the clock frequency $\nu_0$, such that the relative frequency shift remains constant:
		\begin{equation}
			\frac{\delta\nu}{\nu_0} = \frac v c
		\end{equation}
		For this reason, the fountain architecture, for which the Doppler effect is one of the limitations, cannot be applicable to optical clocks, and the atoms have to be tightly confined in a trapping potential to cancel their velocity $v$. To achieve this goal, two different technologies have been developed. First, a single ion is trapped in a RF electric field. These \emph{ion optical clocks}~\cite{Rosenband28032008,PhysRevLett.104.070802,PhysRevLett.108.090801,PhysRevLett.109.203002} achieve record accuracies at $9\times10^{-18}$. However, since only a single ion is trapped ($N=1$, because of the electrostatic repulsion of ions), the stability of theses clocks is limited at the QPN level of $2\times10^{-15}$ at 1~s. Second, a more recent technology involves trapping of a few thousands neutral atoms in a powerful laser standing wave (or optical lattice) by the dipolar force. Due to its power, this trapping potential is highly perturbative, but for a given "\emph{magic}" wavelength of the trapping light~\cite{Katori2005}, the perturbation is equal for the two clock levels, hence cancelled for the clock transition frequency. These \emph{optical lattice clocks}~\cite{Katori2005,le_targat_experimental_2013,PhysRevLett.109.230801,PhysRevLett.103.063001,PhysRevLett.108.183004} have already reached an accuracy of $1\times10^{-16}$ and are rapidly catching up with ion clocks. Furthermore, the large number of interrogated atoms allowed the demonstration of unprecedented stabilities (a few  $1\times10^{-16}$ at 1~s), heading toward their QPN below $1\times10^{-17}$ at 1~s.

	\subsection{Remote comparison of optical frequencies}

The recent breakthrough of performances of optical clocks was permitted by the
development of frequency combs~\cite{ye2004femtosecond}, which realizes a
"ruler" in the frequency domain. These lasers enable the local comparison of
different optical frequencies and comparisons between optical and microwave frequencies. However, although
optical clocks now largely surpass microwave clocks, a complete architecture has
to be established to enable remote comparison of optical frequencies, in order
to validate the accuracy of optical clocks, build "optical" time-scales, or
enable applications of optical clocks such as geodesic measurements
Indeed, the conventional remote frequency comparisons techniques, mainly through GPS links, cannot reach the level of stability and accuracy realized by optical clocks.

%
		\subsubsection*{Fibre links}

		To overcome the limitations of remote clock comparisons using GNSS signals, comparison techniques using optical fibres are being developed. For this, a stable and accurate frequency signal produced by an optical clock is
        sent through a fibre optics that links metrology institutes, directly
        encoded in the phase of the optical carrier. Because of vibrations and
        temperature fluctuations, the fibre adds a significant phase noise to the
        signal. This added noise is measured by comparing the signal after a
        round trip in the fibre to the original signal, and subsequently
        cancelled. Such a signal can be transported through a dedicated fibre
        (dark fibre) when available~\cite{Predehl27042012}, or, more practically
        along with the internet communication (dark channel)~\cite{Lopez:12}.
        These comparison techniques are applicable at a continental scale ; such
        a network will presumably be in operation throughout Europe in the near
        future. In particular, the REFIMEVE+ project will provide a shared
        stable optical oscillator between a large number of laboratories in
        France, with a number of applications beside metrology. This network
        will be connected to international fibre links (NEAT-FT
        project\footnote{\url{http://www.ptb.de/emrp/neatft_home.html}}) that
        will enable long distance comparisons between optical frequency
        standards located in various national metrology laboratories in Europe.
        The ITOC project\footnote{\url{http://projects.npl.co.uk/itoc/}} aims at collecting these comparisons result to demonstrate high accuracy frequency ratios measurements and geophysical applications.

		\subsubsection*{Space-based links}

		When considering inter-continental time and frequency comparisons, only
        satellite link are conceivable. In this aspect, The two-way satellite
        time and frequency transfer (TWSTFT) involves a satellite that actively
        relays a frequency signal in a round-trip configuration. Also, the space
        mission Pharao-ACES~\cite{ACES2009} that involves an ensemble of clocks on board the International Space Station will comprise a number of ground receiver able to remotely compare optical clocks.

	\subsection{Towards a new definition of the SI second based on optical clocks}

        Unlike microwave clocks, for which Cs has been an unquestionable choice
        over half a century, a large number of atomic species seem to be equally
        matched as a new frequency standard based on an optical transition. Ion
        clocks with Hg$^{+}$, Al$^{+}$, Yb$^{+}$, Sr$^{+}$, Ca$^{+}$ have been
        demonstrated, as well as optical lattice clocks with Sr, Yb and more
        prospectively Hg and Mg. As an illustration, four optical transitions
        have already been approved by the CIPM (International Committee for
        Weights and Measures) as secondary representation of the SI second.
        Therefore, although the SI second would already gain in precision with
        optical clock, a clear consensus has yet to emerge before the current
        microwave defined SI second can be replaced.

	\subsection{Applications of optical clocks}

		The level of accuracy reached by optical clocks opens a new range of applications, through the very precise frequency ratios measurements they enable. In fundamental physics, they enable the tracking of dimension-less fundamental constants such as the fine structure constant $\alpha$, the electron to proton mass ratio $\mu = m_e/m_p$, or the quantum chromodynamics mass scale $m_q/\Lambda_\text{QCD}$. Because each clock transition frequency have a different dependence on these constants, their variations imply drifts in clock frequency ratios that can be detected by repeated measurements. Currently, combined optical-to-optical and optical-to-microwave clock comparisons put an upper bound on the relative variation of fundamental constants in the $10^{-17}$/yr range~\cite{Rosenband28032008,PhysRevLett.109.080801,le_targat_experimental_2013,2013arXiv1304.6940L}. Other tests of fundamental physics are also possible with atomic clocks. The local position invariance can be tested by comparing frequency ratios in the course of the earth rotation around the earth~\cite{PhysRevLett.109.080801,le_targat_experimental_2013}, and the gravitational red-shift will be tested with clocks during the Pharao-ACES mission.

		The TAI time scale is created from Cs standards, but recently, the Rb microwave transition started to contribute. In the near future, secondary representations of the SI second based on optical transitions could contribute to TAI, which would thus benefit from their much improved stability.

Clock frequencies, when compared to a coordinate time-scale, are sensitive to
the gravitational potential of the Earth (see section~\ref{sec:relat}).
Therefore it is necessary to take into account their relative height difference
when building atomic time-scales such as TAI. However, the most accurate optical
clocks can resolve a height difference below 10~cm, which is a scale at which
the global gravitational potential is unknown. Because of this, optical clock
could become a tool to precisely measure this potential. They come as a decisive
addition to relative and absolute gravimeters which are sensitive to the
gravitational field, and to satellite based measurement which lack spatial
resolution~(see section~\ref{sec:chrogeo}).



\section{Relativistic frequency transfer}
\label{sec:relat}

In distant comparisons of frequency standards, we are face with the problem of
curvature of space and relative motion of the clocks. These two effects change
locally the flow of proper time with respect to a global coordinate time. In
this section we describe how to compare distant clock frequencies by means of an
electromagnetic signal, and how the comparison is affected by these effects.

\subsection{The Einstein Equivalence Principle}

\begin{figure}[htbp]
   \begin{center}
      \includegraphics[width=\linewidth]{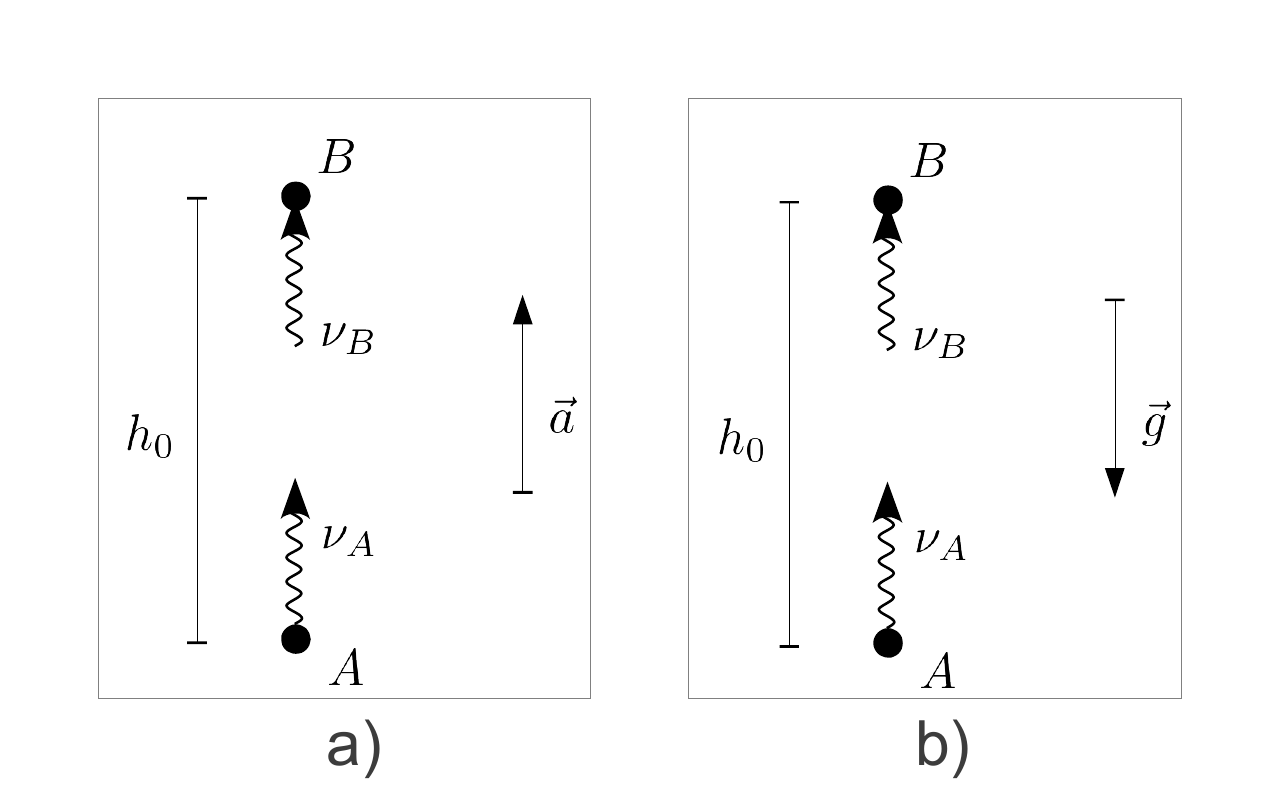}
   \end{center}
   \caption{\label{fig:red} \footnotesize A photon of frequency $\nu_A$ 
   is emitted at point A
   toward point B, where the measured frequency is $\nu_B$. a) $A$ and $B$ are
   two points at rest in an accelerated frame, with acceleration $\vec{a}$ in
   the same direction as the emitted photon. b) $A$ and $B$ are at rest in a non
   accelerated (locally inertial) frame in presence of a gravitational field such
   that $\vec{g} = - \vec{a}$.}
\end{figure}

Before we treat the general case, let's try to understand in simple
terms what is the frequency shift effect. Indeed, it can be seen as a direct
consequence of the Einstein Equivalence Principle~(EEP), one of the pillars of
modern physics~\cite{misner73, will93}. Let's consider a photon emitted at a
point $A$ in an accelerated reference system, toward a point $B$ which lies in
the direction of the acceleration~(see fig.\ref{fig:red}). We assume that both
point are separated by a distance $h_0$, as measured in the accelerated frame. The
photon time of flight is $\delta t = h_0/c$, and the frame velocity during this
time increases by $\delta v = a \delta t = ah_0/c$, where $a$ is the magnitude of
the frame acceleration $\vec{a}$. The frequency at point $B$ (reception) is then
shifted because of Doppler effect, compared to the frequency at point $A$
(emission), by an amount:
\begin{equation}
\frac{\nu_B}{\nu_A} = 1 - \frac{\delta v}{c} = 1 - \frac{ah_0}{c^2}
\end{equation}
Now, the EEP postulates that a gravitational field $\vec{g}$ is locally
equivalent to an acceleration field $\vec{a} = - \vec{g}$. We deduce that in a
non accelerated (locally inertial) frame in presence of a gravitational field $\vec{g}$:
\begin{equation}
\frac{\nu_B}{\nu_A} = 1 - \frac{gh_0}{c^2}
\end{equation}
where $g=|\vec{g}|$, $\nu_A$ is the photon frequency at emission (strong
gravitational potential) and $\nu_B$ is the photon frequency at reception (weak
gravitational field). As $\nu_B<\nu_A$, it is usual to say that the frequency at
the point of reception is ``red-shifted''. One can consider it in terms of
conservation of energy. Intuitively, the photon that goes from $A$ to $B$ has to
``work'' to be able to escape the gravitational field, then it looses energy and
its frequency decreases by virtue of $E= h \nu$, with $h$ the Planck constant.

\begin{figure}[htbp]
   \begin{center}
      \includegraphics[width=\linewidth]{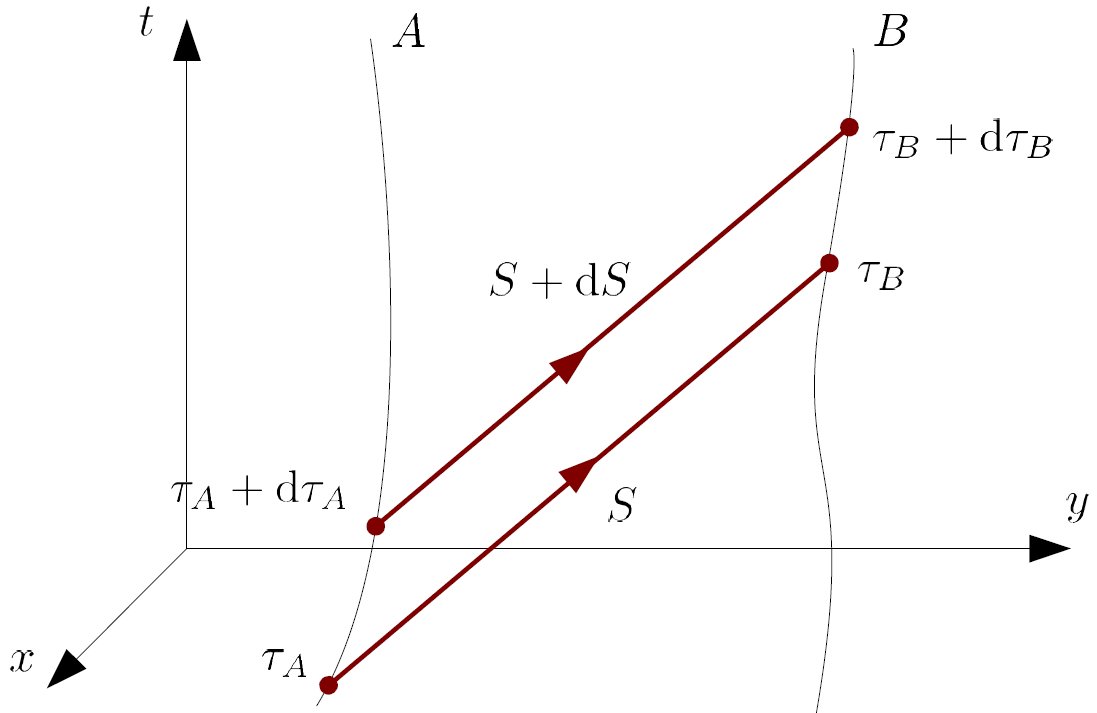}
   \end{center}
   \caption{\label{fig:comp} \footnotesize Two clocks $A$ and $B$ are measuring
   proper time along their trajectory. One signal with phase $S$ is emitted by
   $A$ at proper time $\tau_A$, and another one with phase $S+\dd S$ at time
   $\tau_A + \dd \tau_A$. They are received by clock $B$ respectively at time
   $\tau_B$ and $\tau+\dd \tau_B$.}
\end{figure}

\subsection{General case}
The principle of frequency comparison is to measure the frequency of an
electromagnetic signal with the help of the emitting clock, $A$, and then with
the receiving clock, $B$. We obtain respectively two measurements $\nu_A$ and
$\nu_B$~\footnote{However, in general one measures the time of flight of the
electromagnetic signal between emission and reception. Then the ratio
$\nu_A/\nu_B$ can be obtain by deriving the time of flight measurements.}. Let
$S(x^\alpha)$ be the phase of the electromagnetic signal emitted by clock $A$.
It can be shown that light rays are contained in hypersurfaces of constant
phase. The frequency measured by $A/B$ is:
\begin{equation}
\label{eq:nudef}
\nu_{A/B} = \frac{1}{2\pi} \frac{\dd S}{\dd \tau_{A/B}}
\end{equation}
where $\tau_{A/B}$ is the proper time along the worldline of clock $A/B$ (see
fig.\ref{fig:comp}). We
introduce the wave vector $k^{A/B}_\alpha = (\partial_\alpha S)_{A/B}$ to obtain:
\begin{equation}
\nu_{A/B}= \frac{1}{2\pi} k^{A/B}_\alpha u^\alpha_{A/B}
\end{equation}
where $u^\alpha_{A/B} = \dd x_{A/B}^\alpha / \dd \tau$ is the four-velocity of
clock $A/B$.
Finally, we obtain a fundamental relation for frequency transfer:
\begin{equation}
\dfrac{\nu_A}{\nu_B} = \frac{k^A_\alpha u^\alpha_A}{k^B_\alpha u^\alpha_B}
\end{equation}
This formula does not depend on a particular theory, and then can be used to
perform tests of general relativity. Introducing $v^i = \dd x^i / \dd t$ and
$\hat k_i = k_i / k_0$, it is usually written as:
\begin{equation}
\label{eq:nuk}
\dfrac{\nu_A}{\nu_B} = \frac{u^0_A}{u^0_B} \frac{k_0^A}{k_0^B} \frac{ 1 + \frac{\hat
k^A_i v^i_A}{c} }{ 1 + \frac{\hat k^B_i v^i_B}{c} }
\end{equation}
From eq.~\eqref{eq:nudef} we deduce that:
\begin{equation}
\label{eq:decomp}
\dfrac{\nu_A}{\nu_B} = \frac{\dd \tau_B}{\dd \tau_A}
= \left( \frac{\dd t}{\dd \tau} \right)_A \frac{\dd
t_B}{\dd t_A} \left( \frac{\dd \tau}{\dd t} \right)_B
\end{equation}

We suppose that space-time is stationary, ie. $\partial_0 g_{\alpha \beta} =
0$. Then it can be shown that $k_0$ is constant along the light ray, meaning
that $k_0^A = k_0^B$. We introduce the time transfer function:
\begin{equation}
\TT (x^i_A, x^i_B) = t_B - t_A
\end{equation}
Deriving the time transfer function with respect to $t_A$ one obtains:
\begin{equation}
\label{eq:tTT}
\dfrac{\dd t_B}{\dd t_A} = \dfrac{1 + \frac{\partial \TT}{\partial x^i_A}
v^i_A}{1 - 
\frac{\partial \TT}{\partial x^i_B} v^i_B}
\end{equation}
Inserting eq.~\eqref{eq:tTT} in eq.~\eqref{eq:decomp}, and comparing with
eq.~\eqref{eq:nuk}, we deduce:
\begin{equation}
\hat k_i^A = c \frac{\partial \TT}{\partial x^i_A} \ , \ \hat k_i^B = -c
\frac{\partial \TT}{\partial x^i_B}
\end{equation}
General formula for non-stationary space-times can be found
in~\cite{hees2012a, leponcinlafitte04}.

As an exemple, let's take the simple time transfer function:
\begin{equation}
\TT (x^i_A, x^i_B) = \dfrac{R_{AB}}{c} + \Ol \left( \dfrac{1}{c^3} \right)
\end{equation}
where $R_{AB} = |R_{AB}^i |$ and $R_{AB}^i = x^i_B - x^i_A$. 
Then we obtain:
\begin{equation}
\dfrac{\dd t_B}{\dd t_A} = \dfrac{1 + \frac{\vec{N}_{AB}\cdot \vec{v}_A}{c} +
\Ol \left( \frac{1}{c^3} \right) }{1 + \frac{\vec{N}_{AB}\cdot \vec{v}_B}{c} +
\Ol \left( \frac{1}{c^3} \right) }
\end{equation}
where $N^i_{AB} = R^i_{AB} / R_{AB}$. Up to the second order, this term does not
depend on the gravitational field but on the relative motion of the two clocks.
It is simply the first order Doppler effect. 

The two other terms in eq.~\eqref{eq:decomp} depend on the relation between
proper time and coordinate time. In a metric theory one has $c^2 \dd \tau =
\sqrt{g_{\alpha \beta} \dd x^\alpha \dd x^\beta}$. We deduce that:
\begin{equation}
\label{eq:uu}
\frac{u^0_A}{u^0_B} 
= \dfrac{ \left[ g_{00} + 2 g_{0i} \frac{v^i}{c} + g_{ij}
\frac{v^i v^j}{c^2} \right]^{1/2}_B }{ \left[ g_{00} + 2 g_{0i} \frac{v^i}{c} + g_{ij}
\frac{v^i v^j}{c^2} \right]^{1/2}_A }
\end{equation}

\subsection{Application to a static, spherically symmetric body}

As an example, we apply this formalism for the case of a parametrized post-Minkowskian
approximation of metric theories. This metric is a good approximation of the
space-time metric around the Earth, for a class of theories which is larger than
general relativity. Choosing spatial isotropic coordinates, we assume that
the metric components can be written as:
\begin{align}
g_{00} &= -1+2 (\alpha+1) \frac{GM}{rc^2} 
+ \Ol \left( \frac{1}{c^4} \right)\nonumber\\
g_{0i} &= 0\\
g_{ij} &= \delta_{ij} \left[ 1+2\gamma \frac{GM}{rc^2} 
+ \Ol \left( \frac{1}{c^4} \right) \right]\nonumber
\end{align}
where $\alpha$ and $\gamma$ are two parameters of the theory
(for general relativity $\alpha=0$ and $\gamma=1$). Then:
\begin{equation}
\frac{\dd \tau}{\dd t} = 1 - (\alpha+1) \frac{GM}{rc^2} - \frac{v^2}{2c^2} + \Ol \left( \frac{1}{c^4} \right) 
\end{equation}
where $v = \sqrt{\delta_{ij} v^i v^j}$.
The time transfer function of such a metric is given in eq.~(108) of~\cite{leponcinlafitte04}. Then we calculate:
\begin{equation}
\frac{\dd t_B}{\dd t_A} = \frac{q_A+ \Ol \left( \frac{1}{c^5} \right) }{q_B+ \Ol
\left( \frac{1}{c^5} \right) }
\end{equation}
where
\begin{align}
\label{eq:q1}
q_A =& 1 - \frac{\vec{N}_{AB} \cdot \vec{v}_A}{c} - \frac{2(\gamma + 1)GM}{c^3}
\times \\
&\frac{R_{AB} \vec{N}_A \cdot \vec{v}_A + (r_A + r_B) \vec{N}_{AB} \cdot
\vec{v}_A}{ (r_A+r_B)^2 - R_{AB}^2}\nonumber
\end{align}
\begin{align}
\label{eq:q2}
q_B =& 1 - \frac{\vec{N}_{AB} \cdot \vec{v}_B}{c} - \frac{2(\gamma + 1)GM}{c^3}
\times \\
&\frac{R_{AB} \vec{N}_A \cdot \vec{v}_B - (r_A + r_B) \vec{N}_{AB} \cdot
\vec{v}_B}{ (r_A+r_B)^2 - R_{AB}^2}\nonumber
\end{align}

Let assume that both clocks are at rest with respect to the chosen coordinate
system, ie. $\vec{v}_A = \vec{0} = \vec{v}_B$ and that $r_A = r_0$ and $r_B =
r_0 + \delta r$, where $\delta r \ll r_0$. Then we find:
\begin{equation}
\label{eq:red}
\frac{\nu_A-\nu_B}{\nu_B} = (\alpha+1) \frac{GM}{r_0^2 c^2} \delta r + \Ol
\left( \frac{1}{c^4} \right)
\end{equation}
The same effect can be calculated with a different, not necessarily symmetric
gravitational potential $w(t,x^i)$. The results yields:
\begin{equation}
\label{eq:red2}
\frac{\nu_A-\nu_B}{\nu_B} = \frac{w_A - w_B}{c^2} + \Ol
\left( \frac{1}{c^4} \right)
\end{equation}
This is the classic formula given in textbooks for the ``gravitational
red-shift''. However one should bear in mind that this is
valid for clocks at rest with respect to the coordinate system implicitly
defined by the space-time metric (one uses usually the Geocentric Celestial
Reference System), which is (almost) never the case. Moreover, the
separation between a gravitational red-shift and a Doppler effect is specific to
the approximation scheme used here. One can read the book by
Synge~\cite{synge60} for a different interpretation in terms of relative velocity
and Doppler effect only. 

We note that the lowest order gravitational term in eq.~\eqref{eq:red} is a test
of the Newtonian limit of metric theories.  Indeed, if one wants to recover the
Newtonian law of gravitation for $GM/rc^2 \ll 1$ and $v\ll 1$ then it is
necessary that $\alpha=0$. Then this test is more fundamental than a test of
general relativity, and can be interpreted as a test of Local Position
Invariance (which is a part of the Einstein Equivalence Principle).
See~\cite{will93, will06} for more details on this interpretation, and a review
of the experiments that tested the parameter $\alpha$.

A more realistic case  of the space-time metric is treated in
article~\cite{blanchet01}, in the context of general relativity: all terms from
the Earth gravitational potential are considered up to an accuracy of
$5.10^{-17}$, specifically in prevision of the ACES mission~\cite{ACES2009}.

\section{Chronometric geodesy}
\label{sec:chrogeo}

Instead of using our knowledge of the Earth gravitational field to predict
frequency shifts between distant clocks, one can revert the problem and ask if
the measurement of frequency shifts between distant clocks can improve our
knowledge of the gravitational field. To do simple orders of magnitude
estimates it is good to have in mind some correspondences calculated thanks to
eqs.~\eqref{eq:red} and~\eqref{eq:red2}:
\begin{align}
1 \ \text{meter} \ & \leftrightarrow \frac{\Delta \nu}{\nu} \sim 10^{-16}
\nonumber\\
& \leftrightarrow \Delta w \sim 10 \ \text{m}^2.\text{s}^{-2} \label{eq:corr}
\end{align}
From this correspondence, we can already imagine two direct applications of
clocks in geodesy: if we are capable to compare clocks to $10^{-16}$ accuracy,
we can determine height differences between clocks with one meter accuracy
(levelling), or determine geopotential differences with 10~m$^2$.s$^{-2}$
accuracy.

\subsection{A review of chronometric geodesy}
The first article to explore seriously this possibility was written in 1983 by
Martin Vermeer~\cite{vermeer1983}. The article is named ``chronometric
levelling''. The term ``chronometric'' seems well suited for qualifying the
method of using clocks to determine directly gravitational potential differences, as
``chronometry'' is the science of the measurement of time. However the term
``levelling'' seems too restrictive with respect to all the applications one
could think of using the results of clock comparisons. Therefore we will use the
term ``chronometric geodesy'' to name the scientific discipline that deals with
the measurement and representation of the Earth, including its gravitational
field, with the help of atomic clocks. It is sometimes named ``clock-based
geodesy'', or ``relativistic geodesy''. However this last designation is
improper as relativistic geodesy aims at describing all possible techniques
(including e.g. gravimetry and gradiometry) in a relativistic framework
\cite{kopejkin1991, muller2008, soffel1988}. 

The natural arena of chronometric geodesy is the four-dimensional space-time. At
the lowest order, there is proportionality between relative frequency shift
measurements -- corrected from the first order Doppler effect -- and (Newtonian)
geopotential differences (see eq.\eqref{eq:red2}). To calculate this relation we
have seen that we do not need the theory of general relativity, but only to
postulate Local Position Invariance. Therefore, it is perfectly possible to use
clock comparison measurements -- corrected from the first order Doppler effect
-- as a direct measurement of geopotential differences in the framework of
classical geodesy, if the measurement accuracy does not reach the magnitude of the
higher order terms.

Comparisons between two clocks on the ground generally use a third clock in
space. For the comparison between a clock on the ground and one in space, the
terms of order $c^{-3}$ in eqs.~\eqref{eq:q1}-\eqref{eq:q2} reach a magnitude of
$\sim 10^{-15}$ and $\sim 3 \times 10^{-14}$ for respectively the ground and the
space clock, if they are separated radially by 1000~km. Terms of order $c^{-4}$
omitted in eqs.~\eqref{eq:red}-\eqref{eq:red2} can reach $\sim 5\times
10^{-19}$ in relative frequency shift, which corresponds to a height difference
of $\sim 5$~mm and a geopotential difference of $\sim 5\times
10^{-2}$~m$^2$.s$^{-2}$. Clocks are far from reaching this accuracy today, but
it cannot be excluded for the future.

In his article, Martin Vermeer explores the ``possibilities for technical
realisation of a system for measuring potential differences over
intercontinental distances'' using clock comparisons~\cite{vermeer1983}. The two
main ingredients are of course accurate clocks and a mean to compare them. He
considers hydrogen maser clocks. For the links he considers a 2-way satellite
link over a geostationary satellite, or GPS receivers in interferometric mode.
He has also to consider a mean to compare the proper frequencies of the
different hydrogen maser clocks. However today this can be overcome by comparing
Primary Frequency Standards (PFS), which have a well defined proper frequency
based on a transition of Caesium~133, used for the definition of the second.
However, this problem will rise again if one uses Secondary Frequency Standards
which are not based on Caesium atoms. Then the proper frequency ratio between
two different kinds of atomic clocks has to be determined locally. This is one
of the purpose of the European project ``International timescales with optical
clocks''\footnote{\url{http://projects.npl.co.uk/itoc/}}, where optical clocks
based on different atoms will be compared one each other locally, and to the
PFS. It is planned also to do a proof-of-principle experiment of chronometric
geodesy, by comparing two optical clocks separated by a height difference of
around 1~km using an optical fibre link.

In the foreseen applications of chronometric geodesy, Martin Vermeer mentions
briefly intercontinental levelling, and the measurement of the true geoid
(disentangled from geophysical sea surface effects)~\cite{vermeer1983}. Few
authors have seriously considered chronometric geodesy. Following Vermeer idea,
Brumberg and Groten~\cite{brumberg2002} demonstrated the possibility of using
GPS observations to solve the problem of  the determination of geoid heights,
however leaving aside the practical feasibility of such a technique. Bondarescu
\emph{et al.}~\cite{bondarescu2012} discuss the value and future applicability
of chronometric geodesy, including direct geoid mapping on continents and joint
gravity-geopotential surveying to invert for subsurface density anomalies. They
find that a geoid perturbation caused by a 1.5 km radius sphere with 20 per cent
density anomaly buried at 2 km depth in the Earth's crust is already detectable
by atomic clocks of achievable accuracy. Finally Chou \emph{et al.} demonstrated the
potentiality of the new generation of atomic clocks, based on optical
transitions, to measure heights with a resolution of around
30~cm~\cite{chou2010}.


\subsection{The chronometric geoid}

Arne Bjerhammar in 1985 gives a precise definition of the ``relativistic
geoid''~\cite{bjerhammar1985, bjerhammar1986}:
\begin{quote}
``The relativistic geoid is the surface where precise clocks run with the same
speed and the surface is nearest to mean sea level''
\end{quote}
This is an operational definition. Soffel et al.~\cite{soffel1988} in 1988
translated this definition in the context of post-Newtonian theory. They also
introduce a different operational definition of the relativistic geoid, based
on gravimetric measurements: a surface orthogonal everywhere to the direction of
the plumb-line and closest to mean sea level. He calls the two surfaces obtained
with clocks and gravimetric measurements respectively the ``u-geoid'' and the ``a-geoid''. He proves that these two surfaces coincide in the case of a
stationary metric. In order to distinguish the operational definition
of the geoid from its theoretical description, it is less ambiguous to give a name based
on the particular technique to measure it. Relativistic geoid is too vague as
Soffel et al. have defined two different ones. The names chosen by Soffel et
al. are not particularly explicit, so instead of ``u-geoid'' and ``a-geoid'' one
can call them chronometric and gravimetric geoid respectively. There can be no
confusion with the geoid derived from satellite measurements, as this is a
quasi-geoid that do not coincide with the geoid on the
continents~\cite{hofmann2006}. Other considerations on the chronometric geoid
can be found in~\cite{kopejkin1991, kopeikin2011, muller2008}.

Let two clocks be at rest with respect to the chosen coordinate system ($v^i
= 0$) in an arbitrary space-time. From
formula~\eqref{eq:decomp}, \eqref{eq:tTT} and~\eqref{eq:uu} we deduce that:
\begin{equation}
\frac{\nu_A}{\nu_B} = \frac{\dd \tau_B}{\dd \tau_A} = \frac{\left[ g_{00} \right]_B^{1/2}}{\left[ g_{00}
\right]_A^{1/2}}
\end{equation}
In this case the chronometric geoid is defined by the condition $g_{00} =$~cst.

We notice that the problem of defining a reference surface is closely related to
the problem of realizing Terrestrial Time (TT). TT is defined with respect to
Geocentric Coordinate Time (TCG) by the relation~\cite{soffel03, petit2005}:
\begin{equation}
\frac{\dd TT}{\dd TCG} = 1-L_G
\end{equation}
where $L_G$ is a defining constant. This constant has been chosen so that TT
coincides with the time given by a clock located on the classical geoid. It
could be taken as a formal definition of the chronometric geoid~\cite{wolf1995}.
If so, the chronometric geoid will differ in the future from the classical geoid:
a level surface of the geopotential closest to the topographic mean sea level.
Indeed, the value of the potential on the geoid, $W_0$, depends on the global
ocean level which changes with time~\cite{bursa2007}. With a value of
$\dd W_0 / \dd t \sim 10^{-3}$~m$^2$.s$^{-2}$.y$^{-1}$, the difference in
relative frequency between the classical and the chronometric geoid would be 
around $10^{-18}$ after 100~years (the correspondence is made with the help
of relations~\eqref{eq:corr}). However, the rate of change of the global ocean level could
change during the next decades. Predictions are highly model
dependant~\cite{jevrejeva2012}.

\section{Conclusion}

We presented recent developments in the field of atomic clocks, as well as an
introduction to relativistic frequency transfer and a detailed review of
chronometric geodesy. If the control of systematic effects in optical
clocks keep their promises, they could become very sensible to the gravitational
field, which will ultimately degrade their stability at the surface of the Earth.
One solution will be to send very stable and accurate clocks in space, which
will become the reference against which the Earth clocks would be compared.
Moreover, by sending at least four of these clocks in space, it will be possible
to realize a very stable and accurate four-dimensional reference system in
space~\cite{delva11e}.


\begin{thebibliography}{10}

\bibitem{bjerhammar1985}
A.~{Bjerhammar}.
\newblock {On a relativistic geodesy}.
\newblock {\em Bulletin Geodesique}, 59:207--220, 1985.

\bibitem{bjerhammar1986}
A.~{Bjerhammar}.
\newblock Relativistic geodesy.
\newblock Technical Report NON118 NGS36, NOAA Technical Report, 1986.

\bibitem{blanchet01}
L.~Blanchet, C.~Salomon, P.~Teyssandier, and P.~Wolf.
\newblock Relativistic theory for time and frequency transfer to order
  $c^{-3}$.
\newblock {\em A\&A}, 370:320--329, April 2001.

\bibitem{bondarescu2012}
R.~{Bondarescu}, M.~{Bondarescu}, G.~{Het{\'e}nyi}, L.~{Boschi}, P.~{Jetzer},
  and J.~{Balakrishna}.
\newblock {Geophysical applicability of atomic clocks: direct continental geoid
  mapping}.
\newblock {\em Geophysical Journal International}, 191:78--82, October 2012.

\bibitem{brumberg2002}
V.~A. {Brumberg} and E.~{Groten}.
\newblock {On determination of heights by using terrestrial clocks and GPS
  signals}.
\newblock {\em Journal of Geodesy}, 76:49--54, January 2002.

\bibitem{bursa2007}
M.~{Bur{\v s}a}, S.~{Kenyon}, J.~{Kouba}, Z.~{{\v S}{\'{\i}}ma}, V.~{Vatrt},
  V.~{V{\'{\i}}tek}, and M.~{Vojt{\'{\i}}{\v s}kov{\'a}}.
\newblock {The geopotential value W$_{0}$ for specifying the relativistic
  atomic time scale and a global vertical reference system}.
\newblock {\em Journal of Geodesy}, 81:103--110, February 2007.

\bibitem{ACES2009}
L.~Cacciapuoti and C.~Salomon.
\newblock Space clocks and fundamental tests: The {ACES} experiment.
\newblock {\em The European Physical Journal - Special Topics}, 172:57--68,
  2009.
\newblock 10.1140/epjst/e2009-01041-7.

\bibitem{PhysRevLett.104.070802}
C.~W. Chou, D.~B. Hume, J.~C.~J. Koelemeij, D.~J. Wineland, and T.~Rosenband.
\newblock Frequency comparison of two high-accuracy {Al}$^+$ optical clocks.
\newblock {\em Phys. Rev. Lett.}, 104:070802, Feb 2010.

\bibitem{chou2010}
C.~W. {Chou}, D.~B. {Hume}, T.~{Rosenband}, and D.~J. {Wineland}.
\newblock {Optical Clocks and Relativity}.
\newblock {\em Science}, 329:1630--, September 2010.

\bibitem{delva11e}
P.~Delva, A.~\v{C}ade\v{z}, U.~Kosti\'{c}, and S.~{Carloni}.
\newblock {A relativistic and autonomous navigation satellite system}.
\newblock In {\em Gravitational Waves and Experimental Gravity}, pages
  277--280, 2011.
\newblock Proceedings of the XLVI$^\text{th}$ RENCONTRES DE MORIOND And GPhyS
  Colloquium.

\bibitem{CCDS1980}
P.~{Giacomo}.
\newblock {News from the BIPM}.
\newblock {\em Metrologia}, 17:69--74, April 1981.

\bibitem{6174184}
J.~Gu{\'e}na, M.~Abgrall, D.~Rovera, P.~Laurent, B.~Chupin, M.~Lours,
  G.~Santarelli, P.~Rosenbusch, M.E. Tobar, Ruoxin Li, K.~Gibble, A.~Clairon,
  and S.~Bize.
\newblock Progress in atomic fountains at lne-syrte.
\newblock {\em Ultrasonics, Ferroelectrics and Frequency Control, IEEE
  Transactions on}, 59(3):391--409, 2012.

\bibitem{PhysRevLett.109.080801}
J.~Gu\'ena, M.~Abgrall, D.~Rovera, P.~Rosenbusch, M.~E. Tobar, Ph. Laurent,
  A.~Clairon, and S.~Bize.
\newblock Improved tests of local position invariance using $^{87}\mathrm{Rb}$
  and $^{133}\mathrm{Cs}$ fountains.
\newblock {\em Phys. Rev. Lett.}, 109:080801, Aug 2012.

\bibitem{hees2012a}
A.~{Hees}, S.~{Bertone}, and C.~{Le Poncin-Lafitte}.
\newblock {Frequency shift up to the 2-PM approximation}.
\newblock In S.~{Boissier}, P.~{de Laverny}, N.~{Nardetto}, R.~{Samadi},
  D.~{Valls-Gabaud}, and H.~{Wozniak}, editors, {\em SF2A-2012: Proceedings of
  the Annual meeting of the French Society of Astronomy and Astrophysics},
  pages 145--148, December 2012.

\bibitem{hofmann2006}
Bernhard Hofmann-Wellenhof and Helmut Moritz.
\newblock {\em Physical geodesy}.
\newblock Springer, 2006.

\bibitem{PhysRevLett.108.090801}
N.~Huntemann, M.~Okhapkin, B.~Lipphardt, S.~Weyers, Chr. Tamm, and E.~Peik.
\newblock High-accuracy optical clock based on the octupole transition in
  $^{171}\mathrm{Yb}^{+}$.
\newblock {\em Phys. Rev. Lett.}, 108:090801, Feb 2012.

\bibitem{jevrejeva2012}
S.~{Jevrejeva}, J.~C. {Moore}, and A.~{Grinsted}.
\newblock {Sea level projections to AD2500 with a new generation of climate
  change scenarios}.
\newblock {\em Global and Planetary Change}, 80:14--20, January 2012.

\bibitem{kopejkin1991}
S.~M. {Kopeikin}.
\newblock {Relativistic Manifestations of gravitational fields in gravimetry
  and geodesy}.
\newblock {\em Manuscripta Geodaetica}, 16:301--312, January 1991.

\bibitem{kopeikin2011}
S.~M. {Kopeikin}, M.~Efroimsky, and G.~Kaplan.
\newblock {\em Relativistic celestial mechanics of the solar system}.
\newblock Wiley-VCH, 2011.

\bibitem{leponcinlafitte04}
C.~{Le Poncin-Lafitte}, B.~Linet, and P.~Teyssandier.
\newblock {World function and time transfer: general post-Minkowskian
  expansions}.
\newblock {\em Class. Quantum Grav.}, 21:4463--4483, September 2004.

\bibitem{le_targat_experimental_2013}
R.~Le~Targat, L.~Lorini, Y.~Le~Coq, M.~Zawada, J.~Guéna, M.~Abgrall, M.~Gurov,
  P.~Rosenbusch, D.~G. Rovera, B.~Nagórny, R.~Gartman, P.~G. Westergaard,
  M.~E. Tobar, M.~Lours, G.~Santarelli, A.~Clairon, S.~Bize, P.~Laurent,
  P.~Lemonde, and J.~Lodewyck.
\newblock Experimental realization of an optical second with strontium lattice
  clocks.
\newblock {\em Nat Commun}, 4, July 2013.

\bibitem{2013arXiv1304.6940L}
N.~{Leefer}, C.~T.~M. {Weber}, A.~{Cing{\"o}z}, J.~R. {Torgerson}, and
  D.~{Budker}.
\newblock {New limits on variation of the fine-structure constant using atomic
  dysprosium}.
\newblock {\em ArXiv e-prints}, April 2013.

\bibitem{PhysRevLett.103.063001}
N.~D. Lemke, A.~D. Ludlow, Z.~W. Barber, T.~M. Fortier, S.~A. Diddams,
  Y.~Jiang, S.~R. Jefferts, T.~P. Heavner, T.~E. Parker, and C.~W. Oates.
\newblock Spin-$1/2$ optical lattice clock.
\newblock {\em Phys. Rev. Lett.}, 103:063001, Aug 2009.

\bibitem{Lopez:12}
Olivier Lopez, Adil Haboucha, Bruno Chanteau, Christian Chardonnet, Anne
  Amy-Klein, and Giorgio Santarelli.
\newblock Ultra-stable long distance optical frequency distribution using the
  internet fiber network.
\newblock {\em Opt. Express}, 20(21):23518--23526, Oct 2012.

\bibitem{PhysRevLett.109.203002}
Alan~A. Madej, Pierre Dub\'e, Zichao Zhou, John~E. Bernard, and Marina
  Gertsvolf.
\newblock $^{88}\mathrm{Sr}^{+}$ 445-thz single-ion reference at the
  ${10}^{-17}$ level via control and cancellation of systematic uncertainties
  and its measurement against the si second.
\newblock {\em Phys. Rev. Lett.}, 109:203002, Nov 2012.

\bibitem{PhysRevLett.108.183004}
J.~J. McFerran, L.~Yi, S.~Mejri, S.~Di~Manno, W.~Zhang, J.~Gu\'ena, Y.~Le~Coq,
  and S.~Bize.
\newblock Neutral atom frequency reference in the deep ultraviolet with
  $\mathrm{\text{fractional
  uncertainty}}=5.7\ifmmode\times\else\texttimes\fi{}{10}^{-15}$.
\newblock {\em Phys. Rev. Lett.}, 108:183004, May 2012.

\bibitem{misner73}
C.~W. Misner, K.~S. Thorne, and J.~A. Wheeler.
\newblock {\em {Gravitation}}.
\newblock San Francisco: W.H.~Freeman and Co., 1973, 1973.

\bibitem{muller2008}
J.~{M{\"u}ller}, M.~{Soffel}, and S.~A. {Klioner}.
\newblock {Geodesy and relativity}.
\newblock {\em Journal of Geodesy}, 82:133--145, March 2008.

\bibitem{PhysRevLett.109.230801}
T.~L. Nicholson, M.~J. Martin, J.~R. Williams, B.~J. Bloom, M.~Bishof, M.~D.
  Swallows, S.~L. Campbell, and J.~Ye.
\newblock Comparison of two independent sr optical clocks with
  $1\mathbf{\ifmmode\times\else\texttimes\fi{}}{10}^{-17}$ stability at
  ${10}^{3}\text{\,}\text{\,}\mathbf{s}$.
\newblock {\em Phys. Rev. Lett.}, 109:230801, Dec 2012.

\bibitem{0026-1394-47-1-001}
Thomas~E Parker.
\newblock Long-term comparison of caesium fountain primary frequency standards.
\newblock {\em Metrologia}, 47(1):1, 2010.

\bibitem{6377298}
G.~Petit and G.~Panfilo.
\newblock Comparison of frequency standards used for tai.
\newblock {\em Instrumentation and Measurement, IEEE Transactions on},
  62(6):1550--1555, 2013.

\bibitem{petit2005}
G.~{Petit} and P.~{Wolf}.
\newblock {Relativistic theory for time comparisons: a review}.
\newblock {\em Metrologia}, 42:138, June 2005.

\bibitem{Predehl27042012}
K.~Predehl, G.~Grosche, S.~M.~F. Raupach, S.~Droste, O.~Terra, J.~Alnis, Th.
  Legero, T.~W. Hänsch, Th. Udem, R.~Holzwarth, and H.~Schnatz.
\newblock A 920-kilometer optical fiber link for frequency metrology at the
  19th decimal place.
\newblock {\em Science}, 336(6080):441--444, 2012.

\bibitem{Rosenband28032008}
T.~Rosenband, D.~B. Hume, P.~O. Schmidt, C.~W. Chou, A.~Brusch, L.~Lorini,
  W.~H. Oskay, R.~E. Drullinger, T.~M. Fortier, J.~E. Stalnaker, S.~A. Diddams,
  W.~C. Swann, N.~R. Newbury, W.~M. Itano, D.~J. Wineland, and J.~C. Bergquist.
\newblock Frequency ratio of {Al}$^+$ and {Hg}$^+$ single-ion optical clocks;
  metrology at the 17th decimal place.
\newblock {\em Science}, 319(5871):1808--1812, 2008.

\bibitem{soffel1988}
M.~{Soffel}, H.~{Herold}, H.~{Ruder}, and M.~{Schneider}.
\newblock {Relativistic theory of gravimetric measurements and definition of
  the geoid.}
\newblock {\em Manuscr.~Geod.}, 13:143--146, 1988.

\bibitem{soffel03}
M.~Soffel, S.~A. {Klioner}, G.~{Petit}, P.~{Wolf}, S.~M. {Kopeikin},
  P.~{Bretagnon}, V.~A. {Brumberg}, N.~{Capitaine}, T.~{Damour},
  T.~{Fukushima}, B.~{Guinot}, T.-Y. {Huang}, L.~{Lindegren}, C.~{Ma},
  K.~{Nordtvedt}, J.~C. {Ries}, P.~K. {Seidelmann}, D.~{Vokrouhlick{\'y}},
  C.~M. {Will}, and C.~{Xu}.
\newblock {The IAU 2000 Resolutions for Astrometry, Celestial Mechanics, and
  Metrology in the Relativistic Framework: Explanatory Supplement}.
\newblock {\em The Astronomical Journal}, 126:2687--2706, December 2003.

\bibitem{synge60}
J.~L. Synge.
\newblock {\em {Relativity, The general theory}}.
\newblock North-Holland Publishing Company, Amsterdam, 1960, 1960.

\bibitem{Katori2005}
Masao Takamoto, Feng-Lei Hong, Ryoichi Higashi, and Hidetoshi Katori.
\newblock An optical lattice clock.
\newblock {\em Nature}, 435:321--324, 2005.

\bibitem{0026-1394-4-1-006}
J~Terrien.
\newblock News from the international bureau of weights and measures.
\newblock {\em Metrologia}, 4(1):41, 1968.

\bibitem{vermeer1983}
M.~Vermeer.
\newblock Chronometric levelling.
\newblock {\em {Reports of the Finnish Geodetic Institute}}, 83(2):1--7, 1983.

\bibitem{will93}
C.~M. Will.
\newblock {\em {Theory and Experiment in Gravitational Physics}}.
\newblock Theory and Experiment in Gravitational Physics, by Clifford
  M.\~{}Will, pp.\~{}396.\~{}ISBN 0521439736.\~{}Cambridge, UK: Cambridge
  University Press, March 1993., 1993.

\bibitem{will06}
C.~M. Will.
\newblock {The Confrontation between General Relativity and Experiment}.
\newblock {\em Living Reviews in Relativity}, 9(3), 2006.

\bibitem{wolf1995}
P.~{Wolf} and G.~{Petit}.
\newblock {Relativistic theory for clock syntonization and the realization of
  geocentric coordinate times.}
\newblock {\em A\&A}, 304:653, December 1995.

\bibitem{ye2004femtosecond}
Jun Ye and Steven~T Cundiff.
\newblock {\em Femtosecond optical frequency comb: Principle, Operation and
  Applications}.
\newblock Springer, 2004.

\end{thebibliography}

\end{document}